\documentclass[reprint,aps,pra,showpacs,amsmath,amssymb]{revtex4-1}
\usepackage{dcolumn}
\usepackage{amssymb}
\usepackage{amsmath}
\usepackage{graphicx}
\usepackage{epsfig}
\usepackage{subfigure}
\usepackage{amsfonts}
\usepackage{CJK}
\usepackage{appendix}
\usepackage{multirow}
\usepackage{amsfonts}
\usepackage{dsfont}
\usepackage{amsmath}
\usepackage{graphicx}
\usepackage{float}
\usepackage{amsmath,epsfig,amssymb,subfigure,bm,dsfont}
\usepackage{lipsum}
\usepackage[english]{babel}
\usepackage{bm}
\usepackage{amsmath}
\usepackage[colorlinks,linkcolor=blue,citecolor=blue]{hyperref}%
\usepackage{epstopdf}

\providecommand{\U}[1]{\protect \rule{.1in}{.1in}}
\begin{document}
\title{Bound state in the continuum and dynamics via phase modulation in giant-atom waveguide setups}
\author{Ji Qi$^{1}$}
\author{Xiaojun Zhang$^{1}$}
\author{Hongwei Yu$^{1}$}
\author{Zhihai Wang$^{1}$}
\email{wangzh761@nenu.edu.cn}
\affiliation{1. Center for Quantum Sciences and School of Physics, Northeast Normal University, Changchun 130024, China}

\begin{abstract}
Giant atoms, which couple to a waveguide through multiple spatially separated connection points beyond the dipole approximation, provide a versatile route for quantum information processing based on interference-induced bound states in the continuum (BICs). While multi-giant-atom architectures are being developed toward giant-atom quantum networks, the role of direct coupling between the giant atoms, in particular the associated coupling phase, in atomic dynamics remains insufficiently understood. Here we take a first step toward addressing this issue by studying a two-giant-atom waveguide-QED model. We show that the coupling phase can be used to control both the number of BICs and their profiles for both of photon and atoms. More interestingly, the presence of BICs gives rise to a variety of dynamical behaviors, providing an effective mechanism for tailoring quantum-state evolution in giant-atom waveguide-QED systems. Our results highlight coupling-phase engineering as a useful tool for controlling interference, bound states, and quantum dynamics in nonlocal light--matter interfaces.
\end{abstract}

\maketitle

\section{Introduction}

Bound states in the continuum (BICs) are quantum states embedded in a continuous energy spectrum while remaining spatially localized~\cite{Frank1975,D. C. Marinica2008,M. I. Molina2012,G. Calajo2019,Qiu2023}. Over the past decades, BICs have been extensively investigated in metamaterial and photonic systems, where they have shown great potential for enhancing light--matter interactions~\cite{CW2016,MK2023}. More recently, the investigations about the BIC have been extended to waveguide-QED platforms~\cite{Roy2017,Lehmberg1970a,Lehmberg1970b,Lalumiere2013,Zheng2013,Yin2022,Feng2021}.

In waveguide-QED systems, the waveguide acts as a quantum bus that mediates indirect interactions between distant atoms, with interference effects playing a central role~\cite{D. C. Marinica2008,Qiu2023,Friedrich1985,Fan2003,Weimann2013,Rybin2017,Azzam2018,Leonforte2024,Ingelsten2024}. A paradigmatic example is provided by two spatially separated two-level atoms coupled to a coupled-resonator waveguide (CRW)~\cite{Peng-Bo Li2009}. In this configuration, the two atoms can act as effective atomic mirrors, trapping photons in the intermediate region and thereby supporting BICs in the atom--waveguide composite system~\cite{an2016,Longhi2021,Zhang2023}.

With the experimental realization of strong coupling between transmon qubits~\cite{Kockum2018,Kannan2020,Andersson2019,Vadiraj2021,Gustafsson2014,Guo2017} and surface acoustic waves~\cite{Datta1986,Morgan2007}, the spatial extent of a qubit can become comparable to the wavelength of the propagating field. This breaks the conventional dipole approximation~\cite{Walls2008} and gives rise to the concept of ``giant atoms''~\cite{Andersson2019,Vadiraj2021,Gustafsson2014,Guo2017,Chen2022,L.Du2023,Zhao2020,Kockum2021,Kockum2014,Terradas2022,WangX2021,WangX2022a,Soro2022,LiuN2022,WangX2022b,Joshi2023,Gonzalez2021,Guo2020,GuoS2020,Du2023}. In this emerging paradigm, a giant atom couples nonlocally to a waveguide through multiple spatially separated connection points. As a consequence, photons propagating in the waveguide can undergo multiple reflections between the coupling points, and the resulting interference can induce BICs in both one- and two-dimensional real-space geometries~\cite{Yu2025, soro2023x, ER2024}.

Direct coupling between atoms provides a fundamental quantum resource, such as entanglement, for a wide range of quantum-information-processing tasks. It is therefore natural to ask whether BICs can still emerge, and how they can be controlled, when directly coupled atoms are further coupled to a waveguide. In this work, we address this question in a two-braided-giant-atom waveguide-QED setup. We show that the coupling phase between the two giant atoms acts as an effective control parameter for tuning both the number and spatial profiles of BICs. The resulting phase-engineered BICs can confine photons in markedly different spatial regions and lead to strong atomic entanglement. Moreover, these intrinsic BICs give rise to rich atomic dynamical behaviors, providing a flexible mechanism for controlling quantum dynamics in giant-atom waveguide-QED systems.

The remainder of this paper is organized as follows. In Sec.~\ref{model1}, we introduce the model, where two directly coupled giant atoms are further coupled to a CRW, and investigate the phase-controlled atomic dynamics associated with different numbers of BICs. In Sec.~\ref{profile}, we characterize the profiles of the BICs under the control of the coupling phase by analyzing both the photonic distribution and the atomic-state tomography. Finally, in Sec.~\ref{conclusion}, we give a brief summary.

\begin{figure}[t]
  \centering
  \includegraphics[width=0.5\textwidth]{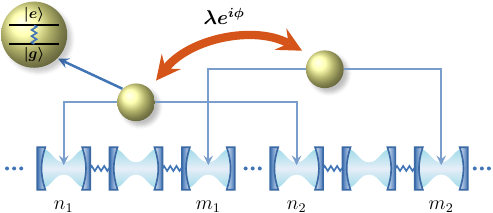}
  \caption{Schematic diagram of two braided giant atoms coupled to a coupled-resonator waveguide (CRW) in Eq.~(\ref{eq5}). In the following sections, we consider that the giant atoms are coupled to the CRW at sites $n_{1}$ and $n_{2}$ ($m_{1}$ and $m_{2}$).}
    \label{fig:model}
\end{figure}

\section{MODEL and dynamics}
\label{model1}

\subsection{Hamiltonian}
As schematically shown in Fig.~\ref{fig:model}, the system under consideration consists of two directly coupled two-level giant atoms. Each giant atom is coupled to a one-dimensional CRW through two spatially separated sites~\cite{Kockum2018}.

The total Hamiltonian can be written as
$H=H_{a}+H_{c}+H_{I}$, where we set $\hbar=1$ throughout. The three terms are given by
\begin{align}
H_{a} &= \Omega\left(|e\rangle_{1}\langle e|+|e\rangle_{2}\langle e|\right)
      + \lambda\left(\sigma_{1}^{+}\sigma_{2}^{-}e^{i\phi}
      + \sigma_{1}^{-}\sigma_{2}^{+}e^{-i\phi}\right), \\
H_{c} &= \omega_{c}\sum_{j} a_{j}^{\dagger}a_{j}
      - \xi\sum_{j}\left(a_{j+1}^{\dagger}a_{j}
      + a_{j}^{\dagger}a_{j+1}\right), \\
H_{I} &= g\left[(a_{n_{1}}+a_{n_{2}})\sigma_{1}^{+}
      + (a_{m_{1}}+a_{m_{2}})\sigma_{2}^{+}\right] \nonumber \\
      &\quad + g\left[(a_{n_{1}}^{\dagger}+a_{n_{2}}^{\dagger})\sigma_{1}^{-}
      + (a_{m_{1}}^{\dagger}+a_{m_{2}}^{\dagger})\sigma_{2}^{-}\right].
\end{align}
Here, $H_{a}$ describes the two directly coupled giant atoms, where $\Omega$ is the transition frequency between the excited state $|e\rangle$ and the ground state $|g\rangle$. The parameters $\lambda$ and $\phi$ denote, respectively, the coupling strength and coupling phase between the two giant atoms.

The term $H_{c}$ describes the CRW, where $\omega_{c}$ is the bare frequency of each identical resonator, $\xi$ is the photon hopping strength between adjacent resonators, and $a_{j}^{\dagger}$ ($a_{j}$) is the creation (annihilation) operator of the $j$th resonator mode.

The interaction Hamiltonian $H_{I}$ characterizes the nonlocal coupling between the giant atoms and the waveguide, with coupling strength $g$. Specifically, the first giant atom is coupled to the resonators at sites $n_{1}$ and $n_{2}$, while the second giant atom is coupled to the resonators at sites $m_{1}$ and $m_{2}$.

In the thermodynamic limit of an infinitely long resonator array, $N_c\to\infty$, we introduce the Fourier transformation $a_j=\sum_k e^{-ikj}a_k/\sqrt{N_c}$ ,
which allows us to rewrite the CRW Hamiltonian in momentum space as~\cite{Longhi2020}
$H_c=\sum_k \omega_k a_k^\dagger a_k$. Here, the dispersion relation is given by
$\omega_k=\omega_c-2\xi\cos k$. Thus, the CRW supports a continuous energy band with bandwidth $4\xi$, centered at $\omega_c$, and thereby provides a structured photonic environment for the giant atoms.

In momentum space, the total Hamiltonian becomes
\begin{equation}
\begin{aligned}
H =&\Omega\left(|e\rangle_{1}\langle e|+|e\rangle_{2}\langle e|\right)
+\sum_k \omega_k a_k^{\dagger}a_k  \\
&+\frac{g}{\sqrt{N_c}}\sum_k
\left[
\left(e^{ikn_{1}}+e^{ikn_{2}}\right)a_k^{\dagger}\sigma_1^{-}
+{\rm H.c.}
\right] \\
&+\frac{g}{\sqrt{N_c}}\sum_k
\left[
\left(e^{ikm_{1}}+e^{ikm_{2}}\right)a_k^{\dagger}\sigma_2^{-}
+{\rm H.c.}
\right] \\
&+\lambda\left(
\sigma_1^{+}\sigma_2^{-}e^{i\phi}
+\sigma_1^{-}\sigma_2^{+}e^{-i\phi}
\right).
\label{eq5}
\end{aligned}
\end{equation}

In this configuration, the waveguide serves as a mediated environment shared by the two giant atoms. Crucially, the coupling phase $\phi$ cannot be eliminated via unitary transformations, which is attributed to the cyclic transitions among the bare states $\sigma_{1}^{+}|G\rangle$, $\sigma_{2}^{+}|G\rangle$, and $a_{k}^{\dagger}|G\rangle$. As demonstrated in the following sections, this phase serves as a functional degree of freedom to tune both the formation of bound states and the atomic dynamics of the system.

\subsection{Amplitudes equations}

Since the rotating-wave approximation has been made for both the atom--waveguide interaction and the photon hopping in the CRW, the total excitation number is conserved. We therefore restrict our discussion to the single-excitation subspace. In this subspace, the state vector can be written as
\begin{equation}
\begin{aligned}
|\psi(t)\rangle =
\left[
\alpha_{1}(t)\sigma_{1}^{+}
+\alpha_{2}(t)\sigma_{2}^{+}
+\sum_{k}\beta_{k}(t)a_{k}^{\dagger}
\right]|G\rangle ,
\end{aligned}
\end{equation}
where $|G\rangle$ denotes the vacuum state in which both giant atoms are in their ground states and all resonators contain no photons. The coefficients $\alpha_{1}(t)$ and $\alpha_{2}(t)$ are the excitation amplitudes of the first and second giant atoms, respectively, while $\beta_{k}(t)$ is the single-photon amplitude of the $k$th waveguide mode.

Substituting the above ansatz into the Schr\"{o}dinger equation,
$i\partial_t|\psi(t)\rangle=H|\psi(t)\rangle$, yields the following coupled equations for the amplitudes

\begin{align}
i\dot{\alpha_{1}}(t) &=\Omega\alpha_{1}(t)+\lambda e^{i\phi}\alpha_{2}(t)\nonumber\\
                     &\quad+\dfrac{g}{\sqrt{N_{c}}}\sum_{k}(e^{-ikn_{1}}+e^{-ikn_{2}})\beta_{k}(t),\label{a1}\\
i\dot{\alpha_{2}}(t) &=\Omega\alpha_{2}(t)+\lambda e^{-i\phi}\alpha_{1}(t)\nonumber \\
                     &\quad+\dfrac{g}{\sqrt{N_{c}}}\sum_{k}(e^{-ikm_{1}}+e^{-ikm_{2}})\beta_{k}(t),\label{a2}\\
i\dot{\beta_{k}}(t)  &=\omega_{k}\beta_{k}(t)+\dfrac{g}{\sqrt{N_{c}}}(e^{ikn_{1}}+e^{ikn_{2}})\alpha_{1}(t)
\nonumber \\
                     &\quad+\dfrac{g}{\sqrt{N_{c}}}
                     (e^{ikm_{1}}+e^{ikm_{2}})\alpha_{2}(t).\label{betak}
\end{align}

Formally solving Eq.~(\ref{betak}) and substituting the solution back into
Eqs.~(\ref{a1}) and~(\ref{a2}), we obtain a closed dynamical equation for the two
giant atoms under the Markov approximation. The resulting dynamics is governed by
\begin{equation}
i\frac{d\vec{\alpha}(t)}{dt}=M\vec{\alpha}(t),
\label{eq:effective_atom_dynamics}
\end{equation}
where $\vec{\alpha}(t)=(\alpha_1(t),\alpha_2(t))^T$. The effective non-Hermitian matrix
$M$ can be written as~\cite{Longhi2021}
\begin{equation}
M=
\begin{pmatrix}
M_{11} & M_{12} \\
M_{21} & M_{22}
\end{pmatrix},
\end{equation}
with
\begin{align}
M_{11}
&=
\Omega-\frac{i g^{2}}{\xi}
\left(1+i^{|n_{1}-n_{2}|}\right),
\\
M_{22}
&=
\Omega-\frac{i g^{2}}{\xi}
\left(1+i^{|m_{1}-m_{2}|}\right),
\\
M_{12}
&=
\lambda e^{i\phi}
-\frac{i g^{2}}{2\xi}
\sum_{p,q=1}^{2}i^{|n_{p}-m_{q}|},
\\
M_{21}
&=
\lambda e^{-i\phi}
-\frac{i g^{2}}{2\xi}
\sum_{p,q=1}^{2}i^{|m_{p}-n_{q}|}.
\end{align}
To obtain the above equations, we have used the formulas \cite{Giuseppe2016}
\begin{equation}
e^{i x \cos k}=\sum_{s=-\infty}^{\infty}i^{s}J_{s}(x),\,
\int_0^\infty d\tau J_m(a\tau)=\frac{1}{|a|},
\end{equation}
where $J_{s}$ denotes the $s$th order Bessel function.

The dynamics of the amplitudes $\alpha_1$ and $\alpha_2$ is then determined by the eigen value of the matrix $M$. Further more, in Ref.~\cite{Longhi2021}, the author stated that the zero imaginary part of the eigen value of $M$ implies the BIC in the atom-waveguide coupled system. Therefore, we first discuss the $M$'s eigen value and leave the profile of the BIC in the next section.

\begin{figure}[t]
  \centering
  \includegraphics[width=0.5\textwidth]{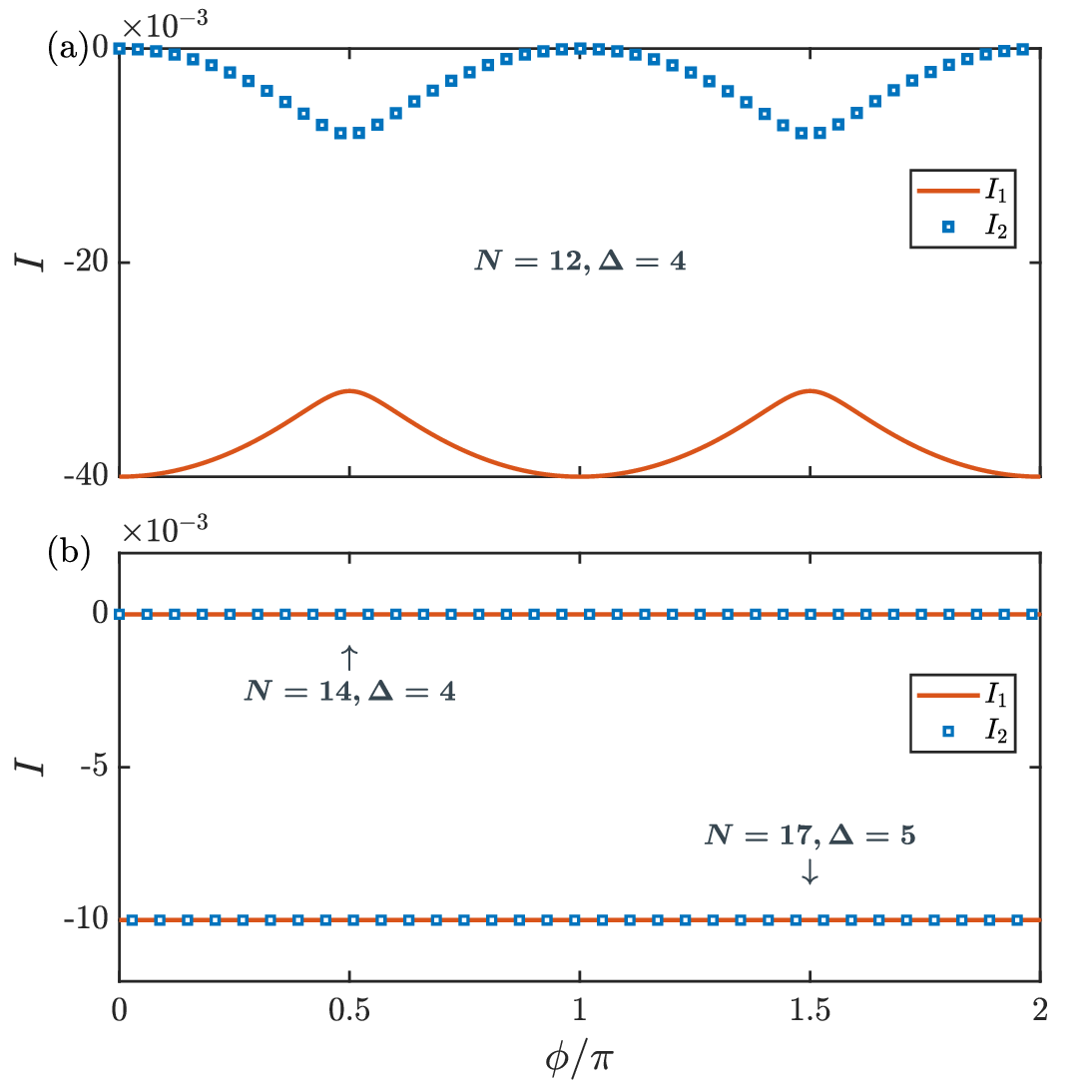}
  \caption{Dependence of the imaginary parts of the eigenvalues of the effective matrix $M$ on the phase $\phi$.(a)$N=12,\Delta=4$, (b)$N=14,\Delta=2$ and $N=17,\Delta=5$. The other parameters are set as $g=0.1\xi, \lambda=1.6 g^{2}/\xi$.}
  \label{fig:giant_atomx}
\end{figure}

We denote the two eigen values of the $M$ as $\lambda_1$ and $\lambda_2$, and plot their imaginary parts $I_i={\rm Im}(\lambda_i),\,i=1,2$ as functions of the coupling phase $\phi$ in Figs.~\ref{fig:giant_atomx} for different configurations. For simplicity, we assume that the two giant atom possess a same size, with $N=n_{2}-n_{1}=m_{2}-m_{1}$ and define the atomic distance as $\Delta=m_1-n_1=m_2-n_2$. In Fig.~\ref{fig:giant_atomx}(a), we observe that both $I_1$ and $I_2$ oscillates periodically with the phase $\phi$ in the setup $N=12,\Delta=4$. Specifically, we find $I_2$ becomes zero as long as $\phi=m\pi,\,m=0,1,2,\cdots$ and $I_1$ is always negative.  According to the statement in Ref.~\cite{Longhi2021}, the system supports one BIC at these phases where $I_2=0$, and no BIC in the regime of $I_1\neq0,I_2\neq0$. In this case, the coupling phase $\phi$ serves as a sensitive controller for the number of the BIC. In Figs.~\ref{fig:giant_atomx}(b), when $N=14,\Delta=4$ we find that there always exist two BICs, i.e., $I_1=I_2=0$, regardless of the phase $\phi$. The similar independence on the phase $\phi$ can be also found in the setup $N=17,\Delta=5$, but $I_1=I_2<0$, implying the complete disappearance of the BICs.

The dynamics of the atomic amplitudes $\alpha_1(t)$ and $\alpha_2(t)$ is determined by the eigenvalues of the effective non-Hermitian matrix $M$. As discussed in Ref.~\cite{Longhi2021}, an eigenvalue of $M$ with a vanishing imaginary part corresponds to a nondecaying component in the atom--waveguide coupled system, and therefore signals the emergence of a BIC. We thus first analyze the eigenvalues of $M$, and then discuss the spatial profiles of the corresponding BICs in the following subsection.

We denote the two eigenvalues of $M$ by $\varepsilon_1$ and $\varepsilon_2$, and plot their imaginary parts, $I_i=\operatorname{Im}(\varepsilon_i),\, i=1,2$, as functions of the coupling phase $\phi$ in Fig.~\ref{fig:giant_atomx} for different geometric configurations. For simplicity, we assume that the two giant atoms have the same size, defined by
$N=n_2-n_1=m_2-m_1$, and introduce the relative separation $\Delta=m_1-n_1=m_2-n_2$.

In Fig.~\ref{fig:giant_atomx}(a), for $N=12$ and $\Delta=4$, both $I_1$ and $I_2$ oscillate periodically with the coupling phase $\phi$. In particular, we find that $I_2$ becomes zero at $\phi=l\pi,\, l=0,1,2,\cdots$, whereas $I_1$ remains negative for all $\phi$. According to the criterion in Ref.~\cite{Longhi2021}, the system therefore supports one BIC at the phases satisfying $I_2=0$, while no BIC exists when both $I_1$ and $I_2$ are nonzero. In this configuration, the coupling phase $\phi$ acts as a sensitive control parameter for tuning the number of BICs.

By contrast, as shown in Fig.~\ref{fig:giant_atomx}(b), for $N=14$ and $\Delta=4$, the two imaginary parts vanish simultaneously, $I_1=I_2=0$, for arbitrary $\phi$. This indicates that the system always supports two BICs, independent of the coupling phase. A similar phase-independent behavior is observed for the configuration $N=17$ and $\Delta=5$, as shown in Fig.~\ref{fig:giant_atomx}(b). However, in this case one has $I_1=I_2<0$ for all $\phi$, implying the complete disappearance of BICs.

\subsection{BIC modulated dynamics}

In the preceding subsection, we have shown that both the geometric configuration and the coupling phase between the giant atoms can serve as sensitive control parameters for tuning the number of BICs in the atom--waveguide coupled system. We now show that these BICs give rise to diverse atomic dynamical behaviors.

\begin{figure}[t]
  \centering
  \includegraphics[width=0.5\textwidth]{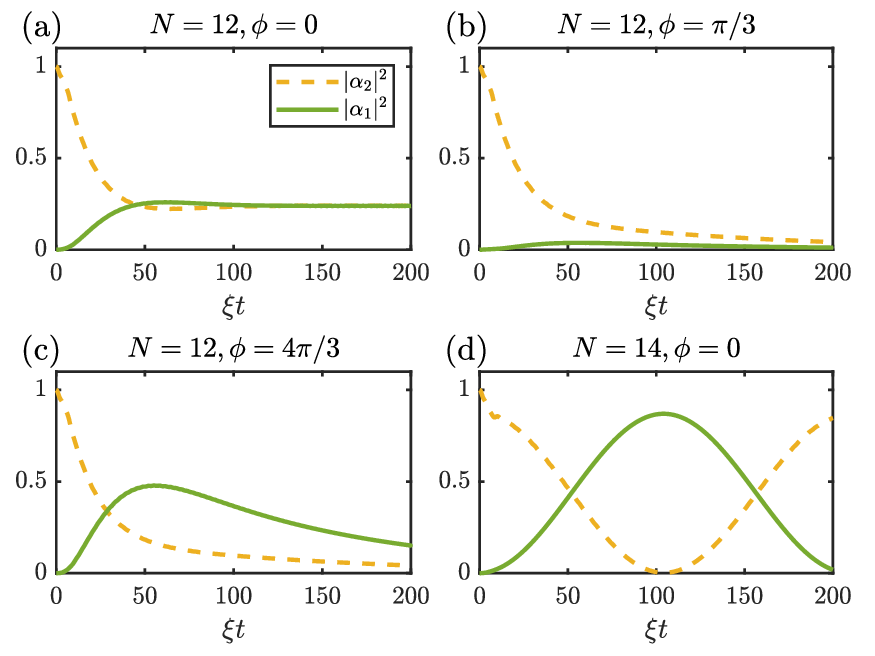}
   \caption{Time evolution of the atomic populations $|\alpha_1(t)|^2$ and $|\alpha_2(t)|^2$.
(a) Dynamics for $N=12$, $\Delta=4$, and $\phi=0$, where the system supports one BIC.
(b,c) Dynamics for the same geometry but at different coupling phases where no BIC is supported.
(d) Dynamics for $N=14$, $\Delta=4$, and $\phi=0$, where the system supports two BICs and exhibits long-lived Rabi-like oscillations. The other parameters are set as $g=0.1\xi, \lambda=1.6 g^{2}/\xi$.}
  \label{fig:giant_atom}
\end{figure}

To this end, we consider the system initially prepared in the state $|\psi(0)\rangle=\sigma_1^+|G\rangle$, where $|G\rangle$ denotes the state in which both giant atoms are in their ground states and the waveguide is in the vacuum state. In Fig.~\ref{fig:giant_atom}, we calculate the atomic populations $|\alpha_{1}(t)|^{2}$ and $|\alpha_{2}(t)|^{2}$ to characterize the time evolution of the two giant atoms.

First, we consider the case in which the system supports a single BIC, corresponding to the configuration shown in Fig.~\ref{fig:giant_atomx}(a), and choose the coupling phase as $\phi=0$. As shown in Fig.~\ref{fig:giant_atom}(a), the initially excited giant atom, i.e., atom $1$, relaxes to a finite nonzero steady-state population. Meanwhile, the initially unexcited giant atom, i.e., atom $2$, also acquires the same long-time population. This behavior can be understood from the formation of a BIC, which prevents complete radiative decay into the waveguide modes and leaves a fractional atomic population trapped in the atom--waveguide coupled system.

This is in sharp contrast to the cases shown in Figs.~\ref{fig:giant_atom}(b) and~\ref{fig:giant_atom}(c), where the BIC is absent due to the choice of $\phi\neq l\pi$ with integer $l$. In these cases, both atoms eventually decay to their ground states, although the initially unexcited atom may transiently acquire a finite excitation during the evolution. Furthermore, even in the absence of BICs, the coupling phase can still be used to modulate the atomic dynamics. This is clearly reflected by the markedly different transient behaviors in Figs.~\ref{fig:giant_atom}(b) and~\ref{fig:giant_atom}(c), which correspond to $\phi=\pi/3$ and $\phi=4\pi/3$, respectively, for the same atomic geometry. Similar complete-dissipation dynamics is also observed for the configuration with $N=17$ and $\Delta=5$, where no BIC exists, and is therefore not shown here.

We next consider the case in which two BICs are present. For the configuration shown in Fig.~\ref{fig:giant_atomx}(b), with $N=14$ and $\Delta=4$, the atomic dynamics is independent of the coupling phase. As shown in Fig.~\ref{fig:giant_atom}(d), the population of the initially excited atom first undergoes a rapid decay over a short time scale, and then persistently exchanges excitation with the second atom through long-lived Rabi-like oscillations. Similar oscillatory dynamics has also been reported in two-giant-atom systems without direct coupling~\cite{Du2023,soro2023x,Yu2025}.

\section{Profile of BIC}
\label{profile}

In the previous section, we showed that the existence of BICs can be exploited to control the atomic dynamics effectively. We now further characterize these BICs by examining their spatial profiles, including both the photonic distribution and the atomic-state structure, and show how they can be modulated by the atomic coupling phase $\phi$.

\begin{figure}[t]
  \centering
  \includegraphics[width=0.5\textwidth]{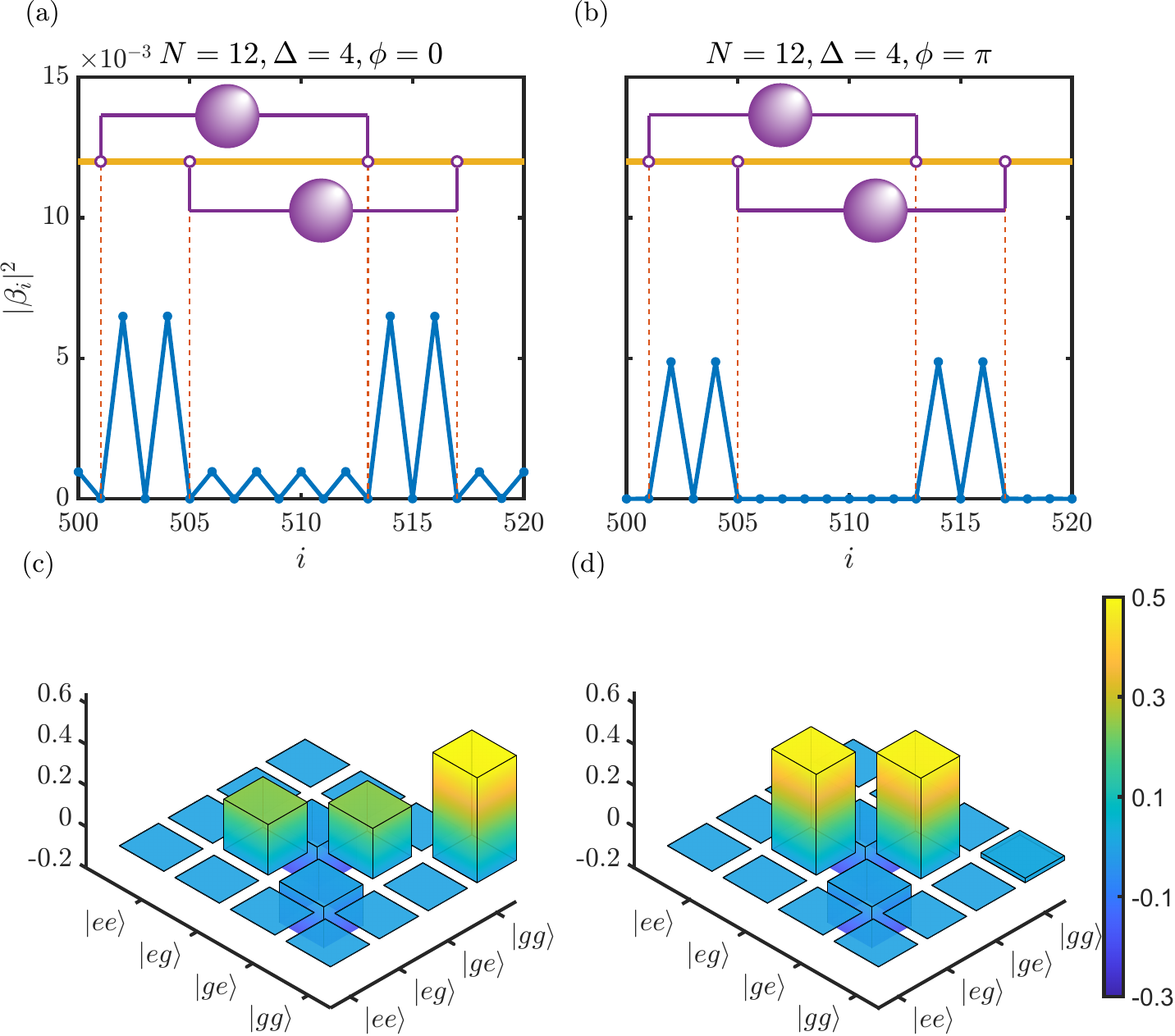}
\caption{Profiles of the BICs and the corresponding atomic-state tomography.
(a,b) Spatial profiles of the BICs for $N=12$ and $\Delta=4$ at coupling phases $\phi=0$ and $\phi=\pi$, respectively.
(c,d) Density matrices of the corresponding atomic states associated with the BICs shown in (a) and (b), respectively. The other parameters are set as $g=0.1\xi, \lambda=1.6 g^{2}/\xi$.}
 \label{fig:giant_atom1}
\end{figure}

An arbitrary eigenstate of the atom--waveguide coupled system satisfies the stationary Schr\"odinger equation $H|E\rangle = E|E\rangle$, where, within the single-excitation subspace, the eigenstate can be written as $|E\rangle =\left[\alpha_1 \sigma_1^++\alpha_2 \sigma_2^+
+\sum_i \beta_i a_i^\dagger\right]|G\rangle$. Here, $\alpha_1$ and $\alpha_2$ denote the excitation amplitudes of the two giant atoms, while $\beta_i$ is the photonic amplitude at the $i$th resonator site. In Fig.~\ref{fig:giant_atom1}, we plot the photonic distribution $|\beta_i|^2$ together with the tomography of the atomic state for the BICs, thereby illustrating their photonic and atomic profiles.

We first consider the case in which only one BIC is present, as shown in Fig.~\ref{fig:giant_atomx}(a) for $N=12$, $\Delta=4$, and $\phi=m\pi$ with $m\in\mathbb{Z}$. The results in Figs.~\ref{fig:giant_atom1}(a) and~\ref{fig:giant_atom1}(b) show that the supported BICs exhibit qualitatively similar spatial profiles for $\phi=0$ and $\phi=\pi$. In both cases, the photon is mainly confined between the $n_1$th and $m_1$th sites, as well as between the $n_2$th and $m_2$th sites; that is, between the left coupling points of the two giant atoms and between their right coupling points. The photonic distribution $|\beta_i|^2$ reaches its maxima when $i-n_1$ is odd, while it vanishes when $i-n_1$ is even. Nevertheless, the nonzero photonic amplitudes for $\phi=0$ are relatively larger than those for $\phi=\pi$. In addition, for $\phi=0$, a weak photonic excitation can still be observed in the region between the right coupling point of the first giant atom and the left coupling point of the second giant atom. By contrast, for $\phi=\pi$, the photonic amplitude in this intermediate region vanishes completely.

The corresponding atomic states of the BICs are shown by state tomography in Figs.~\ref{fig:giant_atom1}(c) and~\ref{fig:giant_atom1}(d) for $\phi=0$ and $\phi=\pi$, respectively. For $\phi=0$, the reduced atomic state has a considerable population in the ground state, while the remaining population lies in the single-excitation subspace. In contrast, for $\phi=\pi$, the two atoms are almost populated in a Bell-like state. Consequently, the atoms become strongly entangled, with concurrence $\mathcal{C}=0.98$ for $\phi=\pi$, which is markedly different from the value $\mathcal{C}=0.49$ obtained for $\phi=0$.

\begin{figure}[t]
  \centering
  \includegraphics[width=0.5\textwidth]{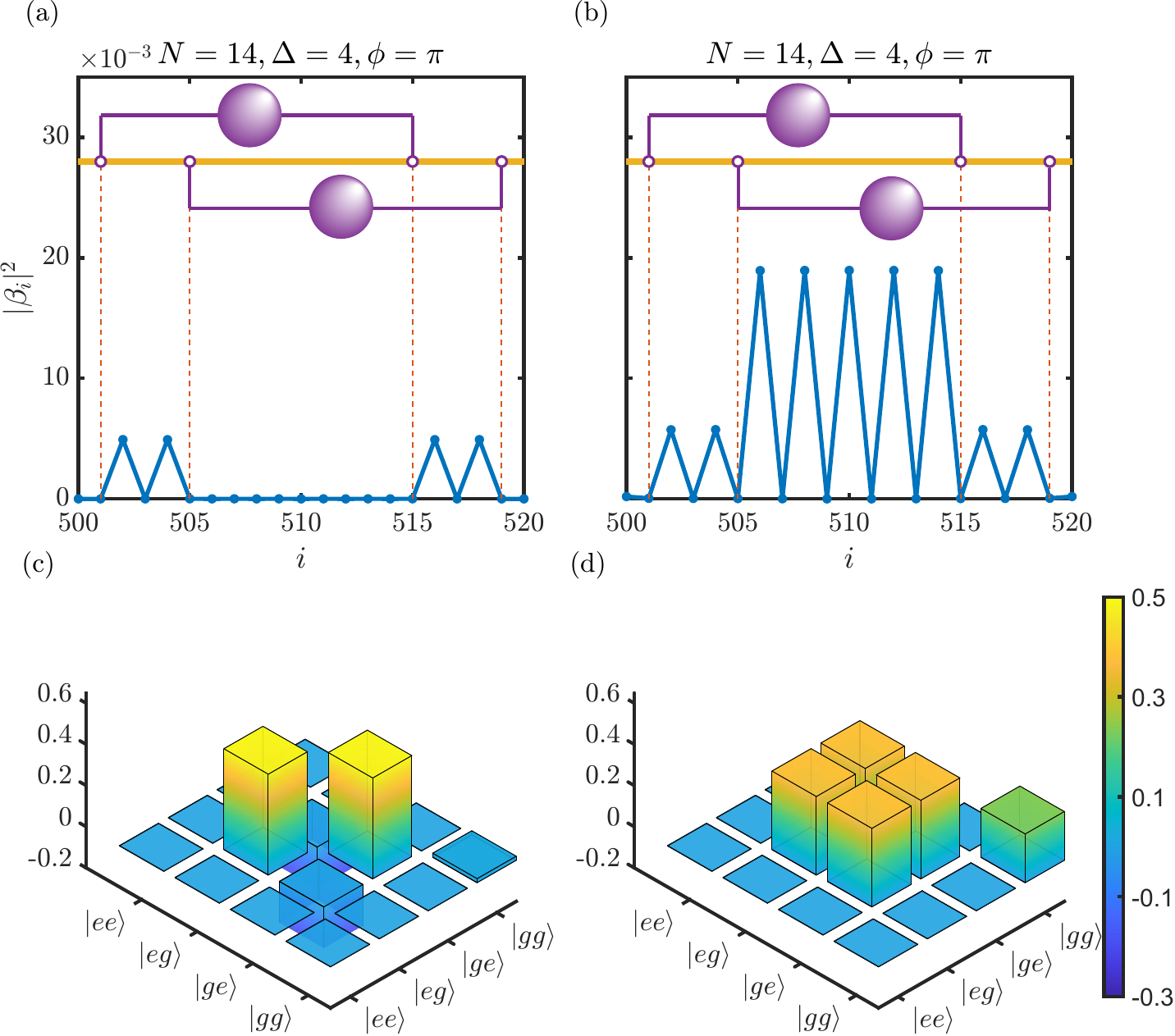}
  \caption{Profiles of the BICs and the corresponding atomic-state tomography. For $N=14$ and $\Delta=4$, the system supports two BICs.
(a,b) Spatial profiles of the two BICs for $\phi=\pi$.
(c,d) Density matrices of the corresponding atomic states for the BICs shown in (a) and (b), respectively. The other parameters are set as $g=0.1\xi, \lambda=1.6 g^{2}/\xi$.} % ?       ?
  \label{fig:giant_atom2}
\end{figure}

Next, we consider the case with two BICs, corresponding to the configuration $N=14$ and $\Delta=4$. In this setup, both the eigenvalues of $M$ and the atomic dynamics are insensitive to the coupling phase $\phi$. One of the two BICs is shown in Fig.~\ref{fig:giant_atom2}(a), which exhibits a spatial profile similar to that of the BICs in the configuration $N=12$ and $\Delta=4$, as shown in Figs.~\ref{fig:giant_atom1}(a) and~\ref{fig:giant_atom1}(b). For this BIC, the corresponding atomic-state tomography is presented in Fig.~\ref{fig:giant_atom2}(c). The atomic component is mainly distributed in the single-excitation subspace, giving rise to strong atomic entanglement with concurrence $\mathcal{C}=0.98$.

The other BIC, shown in Fig.~\ref{fig:giant_atom2}(b), confines the photon mainly in the region between the right coupling point of the first giant atom and the left coupling point of the second giant atom. In the remaining region covered by the two giant atoms, the photonic excitation has only a small amplitude. As shown by the atomic-state tomography in Fig.~\ref{fig:giant_atom2}(d), the admixture of the atomic ground state reduces the degree of entanglement, and the concurrence decreases to $\mathcal{C}=0.76$.

\section{CONCLUSION}
\label{conclusion}

In this work, we have systematically investigated interference effects in waveguide-coupled giant-atom systems, with particular emphasis on the role of the atomic coupling phase in the formation of BICs and the resulting atomic dynamics. Specifically, we consider a pair of directly coupled giant atoms arranged in a braided geometry. By tuning the atomic coupling phase, the system can be engineered to support two, one, or no BICs. The resulting BICs can confine photons in different spatial regions and generate entanglement between the two atoms. These phase-controllable BICs further give rise to diverse atomic dynamical behaviors, including fractional population trapping, complete radiative decay, and long-lived Rabi-like oscillations.

Our proposal is experimentally relevant to superconducting quantum circuits, where giant atoms can be implemented using transmon qubits coupled to a waveguide composed of multiple resonator sites. For experimentally achievable photon hopping strengths in the range $\xi\in(100,200)\,{\rm MHz}$~\cite{P.Roushan2017}, the dynamical time scale demonstrated in Fig.~3 is on the order of several microseconds, which is shorter than the typical decoherence time of transmon qubits, $T_1\approx 10\,{\rm \mu s}$~\cite{Kockum2018,Gustafsson2014}. These estimates suggest that the predicted phase-controlled BIC formation and the associated atomic dynamics should be observable with current superconducting-circuit technologies.

Our results establish a unified physical picture of phase-controlled interference and bound-state formation in giant-atom waveguide-QED systems. They provide a versatile framework for engineering dissipation-free states and tailoring dynamical properties in open quantum systems, with potential applications in quantum networks, quantum state storage, and giant-atom-based quantum devices.

\section*{Acknowledgments}

This work is supported by the the Quantum Science and Technology-National Science and
Technology Major Project (No. 2023ZD0300700) and Natural Science Foundation of China (Grants Nos.~12375010).

\section*{DATA AVAILABILITY}

The data that support the findings of this article are not publicly available. The data are available from the authors upon reasonable request.


\begin{thebibliography}{99}

\bibitem{Frank1975} F. H. Stillinger and D. R. Herrick, Bound states in the continuum, Phys. Rev. A \textbf{11}, 446 (1975).

\bibitem{M. I. Molina2012}M. I. Molina, A. E. Miroshnichenko, and Y. S. Kivshar, Surface bound states in the continuum, Phys. Rev. Lett. \textbf{108}, 070401 (2012).

\bibitem{G. Calajo2019}G. Calaj\`o, Y. L. Fang, H. U. Baranger, and F. Ciccarello, Exciting a bound state in the continuum through multiphoton scattering plus delayed quantum feedback, Phys. Rev. Lett. \textbf{122}, 073601 (2019).

\bibitem{Qiu2023}Q.-Y. Qiu, Y. Wu, and X.-Y. L\"{u}, Collective radiance of giant atoms in non-Markovian regime, Sci. China Phys. Mech. Astron. \textbf{66}, 224212 (2023).

\bibitem{D. C. Marinica2008}D. C. Marinica, A. G. Borisov, and S. V. Shabanov, Bound states in the continuum in photonics, Phys. Rev. Lett. \textbf{100}, 183902 (2008).

\bibitem{CW2016} C. W. Hsu, B. Zhen, A. D. Stone, J. D. Joannopoulos, and M. Solja\v{c}i\'{c}, Bound states in the continuum, Nat. Rev. Mater. \textbf{1}, 16048 (2016).

\bibitem{MK2023} M. Kang, T. Liu, C. T. Chan, and M. Xiao, Applications of bound states in the continuum in photonics, Nat. Rev. Phys. \textbf{5}, 659 (2023).
%waveguige%

\bibitem{Roy2017}D. Roy, C. M. Wilson, and O. Firstenberg, Colloquium: Strongly interacting photons in one-dimensional continuum, Rev. Mod. Phys. \textbf{89}, 021001 (2017).



\bibitem{Lehmberg1970a}R. H. Lehmberg, Radiation from an $N$-atom system. I. General formalism, Phys. Rev. A \textbf{2}, 883 (1970).



\bibitem{Lehmberg1970b}R. H. Lehmberg, Radiation from an $N$-atom system. II. Spontaneous emission from a pair of atoms, Phys. Rev. A \textbf{2}, 889 (1970).



\bibitem{Lalumiere2013}K. Lalumi\`ere, B. C. Sanders, A. F. van Loo, A. Fedorov, A. Wallraff, and A. Blais, Input-output theory for waveguide QED with an ensemble of inhomogeneous atoms, Phys. Rev. A \textbf{88}, 043806 (2013).



\bibitem{Zheng2013}H. Zheng and H. U. Baranger, Persistent quantum beats and long-distance entanglement from waveguide-mediated interactions, Phys. Rev. Lett. \textbf{110}, 113601 (2013).



\bibitem{Yin2022}X.-L. Yin, W.-B. Luo, and J.-Q. Liao, Non-Markovian disentanglement dynamics in double-giant-atom waveguide-QED systems, Phys. Rev. A \textbf{106}, 063703 (2022).


\bibitem{Feng2021}S. L. Feng and W. Z. Jia, Manipulating single-photon transport in a waveguide-QED structure containing two giant atoms, Phys. Rev. A \textbf{104}, 063712 (2021).



%CRW%





%干涉效应%

\bibitem{Friedrich1985}H. Friedrich and D. Wintgen, Interfering resonances and bound states in the continuum, Phys. Rev. A \textbf{32}, 3231 (1985).



\bibitem{Fan2003}S. Fan, W. Suh, and J. D. Joannopoulos, Temporal coupled-mode theory for the Fano resonance in optical resonators, J. Opt. Soc. Am. A \textbf{20}, 569 (2003).



\bibitem{Weimann2013}S. Weimann, Y. Xu, R. Keil, A. E. Miroshnichenko, A. T\"{u}nnermann, S. Nolte, A. A. Sukhorukov, A. Szameit, and Y. S. Kivshar, Compact surface Fano states embedded in the continuum of waveguide arrays, Phys. Rev. Lett. \textbf{111}, 240403 (2013).



\bibitem{Rybin2017}M. V. Rybin, K. L. Koshelev, Z. F. Sadrieva, K. B. Samusev, A. A. Bogdanov, M. F. Limonov, and Y. S. Kivshar, High-$Q$ supercavity modes in subwavelength dielectric resonators, Phys. Rev. Lett. \textbf{119}, 243901 (2017).



\bibitem{Azzam2018}S. I. Azzam, V. M. Shalaev, A. Boltasseva, and A. V. Kildishev, Formation of bound states in the continuum in hybrid plasmonic-photonic systems, Phys. Rev. Lett. \textbf{121}, 253901 (2018).



\bibitem{Leonforte2024}L. Leonforte, X. Sun, D. Valenti, B. Spagnolo, F. Illuminati, A. Carollo, and F. Ciccarello, Quantum optics with giant atoms in a structured photonic bath, Quantum Sci. Technol. \textbf{10}, 015057 (2024).



\bibitem{Ingelsten2024}E. R. Ingelsten, A. F. Kockum, and A. Soro, Avoiding decoherence with giant atoms in a two-dimensional structured environment, Phys. Rev. Res. \textbf{6}, 043222 (2024).



\bibitem{Peng-Bo Li2009}P.-B. Li, Y. Gu, Q.-H. Gong, and G.-C. Guo, Quantum-information transfer in a coupled resonator waveguide, Phys. Rev. A \textbf{79}, 042339 (2009).


%超导量子比特%

\bibitem{an2016}C. Chen, C. Yang, and J. An, Exact decoherence-free state of two distant quantum systems in a non-Markovian environment, Phys. Rev. A \textbf{93}, 062122 (2016).

\bibitem{Longhi2021}S. Longhi, Rabi oscillations of bound states in the continuum, Opt. Lett. \textbf{46}, 2091 (2021).

\bibitem{Zhang2023}X. Zhang, C. Liu, Z. Gong, and Z. Wang, Quantum interference and controllable magic cavity QED via a giant atom in a coupled resonator waveguide, Phys. Rev. A \textbf{108}, 013704 (2023).

\bibitem{Kockum2018}A. F. Kockum, G. Johansson, and F. Nori, Decoherence-free interaction between giant atoms in waveguide quantum electrodynamics, Phys. Rev. Lett. \textbf{120}, 140404 (2018).



\bibitem{Kannan2020}B. Kannan, M. J. Ruckriegel, D. L. Campbell, A. F. Kockum, J. Braum\"{u}ller, D. K. Kim, M. Kjaergaard, P. Krantz, A. Melville, B. M. Niedzielski, A. Veps\"{a}l\"{a}inen, R. Winik, J. L. Yoder, F. Nori, T. P. Orlando, S. Gustavsson, and W. D. Oliver, Waveguide quantum electrodynamics with superconducting artifcial giant atoms, Nature (London) \textbf{583}, 775 (2020).



\bibitem{Andersson2019}G. Andersson, B. Suri, L. Guo, T. Aref, and P. Delsing, Non-exponential decay of a giant artificial atom, Nature Phys. \textbf{15}, 1123 (2019).



\bibitem{Vadiraj2021}A. M. Vadiraj, A. Ask, T. G. McConkey, I. Nsanzineza, C. W. Sandbo Chang, A. F. Kockum, and C. M. Wilson, Engineering the level structure of a giant artificial atom in waveguide quantum electrodynamics, Phys. Rev. A \textbf{103}, 023710 (2021).

\bibitem{Gustafsson2014}M. V. Gustafsson, T. Aref, A. F. Kockum, M. K. Ekstr\"{o}m, G. Johansson, and P. Delsing, Propagating phonons coupled to an artificial atom, Science \textbf{346}, 207 (2014).



\bibitem{Guo2017}L. Guo, A. Grimsmo, A. F. Kockum, M. Pletyukhov, and G. Johansson, Giant acoustic atom: A single quantum system with a deterministic time delay, Phys. Rev. A \textbf{95}, 053821 (2017).



%表面声波%

\bibitem{Datta1986}S. Datta, \textit{Surface Acoustic Wave Devices} (Prentice-Hall, Englewood Cliffs, 1986).



\bibitem{Morgan2007}D. Morgan, \textit{Surface Acoustic Wave Filters}, 2nd ed. (Academic, Amsterdam, 2007).



%电偶极近似%

\bibitem{Walls2008}D. F. Walls and G. J. Milburn, \textit{Quantum Optics}, 2nd ed. (Springer, Berlin, 2008).





%giant atom%

\bibitem{Chen2022}Y. Chen, L. Du, L. Guo, Z. Wang, Y. Zhang, Y. Li, and J. Wu, Nonreciprocal and chiral single-photon scattering for giant atoms, Commun. Phys. \textbf{5}, 215 (2022).

\bibitem{L.Du2023}L. Du, Y. Zhang, and Y. Li, A giant atom with modulated transition frequency, Front. Phys. \textbf{18}, 12301 (2023).

\bibitem{Zhao2020}W. Zhao and Z. Wang, Single-photon scattering and bound states in an atom-waveguide system with two or multiple coupling points, Phys. Rev. A \textbf{101}, 053855 (2020).

\bibitem{Kockum2021}A. F. Kockum, Quantum optics with giant atoms---the first five years, in \textit{International Symposium on Mathematics, Quantum Theory, and Cryptography}, edited by T. Takagi \textit{et al.} (Springer, Singapore, 2021), p. 125.

\bibitem{Kockum2014}A. F. Kockum, P. Delsing, and G. Johansson, Designing frequency-dependent relaxation rates and Lamb shifts for a giant artificial atom, Phys. Rev. A \textbf{90}, 013837 (2014).

\bibitem{Terradas2022}S. Terradas-Brians\'{o}, C. A. Gonz\'{a}lez-Guti\'{e}rrez, F. Nori, L. Mart\'{i}n-Moreno, and D. Zueco, Ultrastrong waveguide QED with giant atoms, Phys. Rev. A \textbf{106}, 063717 (2022).

\bibitem{WangX2021}X. Wang, T. Liu, A. F. Kockum, H.-R. Li, and F. Nori, Tunable chiral bound states with giant atoms, Phys. Rev. Lett. \textbf{126}, 043602 (2021).

\bibitem{WangX2022a}X. Wang, Z. Gao, J. Li, H. Zhu, and H.-R. Li, Unconventional quantum electrodynamics with a Hofstadter ladder waveguide, Phys. Rev. A \textbf{106}, 043703 (2022).

\bibitem{Soro2022}A. Soro and A. F. Kockum, Chiral quantum optics with giant atoms, Phys. Rev. A \textbf{105}, 023712 (2022).

\bibitem{LiuN2022}N. Liu, X. Wang, X. Wang, X.-S. Ma, and M.-T. Cheng, Tunable single photon nonreciprocal scattering based on giant atom-waveguide chiral couplings, Opt. Express \textbf{30}, 23428 (2022).

\bibitem{WangX2022b}X. Wang and H. R. Li, Chiral quantum network with giant atoms, Quantum Sci. Technol. \textbf{7}, 035007 (2022).

\bibitem{Joshi2023}C. Joshi, F. Yang, and M. Mirhosseini, Resonance fluorescence of a chiral artificial atom, Phys. Rev. X \textbf{13}, 021039 (2023).

\bibitem{Gonzalez2021}C. A. Gonz\'{a}lez-Guti\'{e}rrez, J. Rom\'{a}n-Roche, and D. Zueco, Distant emitters in ultrastrong waveguide QED: Ground-state properties and non-Markovian dynamics, Phys. Rev. A \textbf{104}, 053701 (2021).

\bibitem{Guo2020}L. Guo, A. F. Kockum, F. Marquardt, and G. Johansson, Oscillating bound states for a giant atom, Phys. Rev. Res. \textbf{2}, 043014 (2020).

\bibitem{GuoS2020}S. Guo, Y. Wang, T. Purdy, and J. Taylor, Beyond spontaneous emission: Giant atom bounded in the continuum, Phys. Rev. A \textbf{102}, 033706 (2020).

\bibitem{Du2023}L. Du, L. Guo, and Y. Li, Complex decoherence-free interactions between giant atoms, Phys. Rev. A \textbf{107}, 023705 (2023).

%结论1%

\bibitem{soro2023x}A. Soro, C. S. Mu\~{n}oz, and A. F. Kockum, Phys. Rev. A \textbf{107}, 013710 (2023).

\bibitem{Yu2025}H. Yu, X. Zhang, Z. Wang, and J. Wang, Rabi oscillation and fractional population via the bound states in the continuum in a giant atom waveguide QED setup, Phys. Rev. A \textbf{111}, 053710 (2025).

\bibitem{ER2024}E. R. Ingelsten, A. F. Kockum, and A. Soro, Avoiding decoherence with giant atoms in a two-dimensional structured environment, Phys. Rev. Res. \textbf{6}, 043222 (2024).



%oscillation%



%fractional populations%






%%







\bibitem{Longhi2020}S. Longhi, Photonic simulation of giant atom decay, Opt. Lett. \textbf{45}, 3017 (2020).

\bibitem{Giuseppe2016}G. Calaj\'o, F. Ciccarello, D. Chang, and P. Rabl, Atom-field dressed states in slow-light waveguide QED, Phys. Rev. A \textbf{93}, 033833 (2016).



%巨原子实验%



\bibitem{P.Roushan2017}P. Roushan, C. Neill, J. Tangpanitanon, V. M. Bastidas, A. Megrant, R. Barends, Y. Chen, Z. Chen, B. Chiaro, A. Dunsworth, et al., Spectroscopic signatures of localization with interacting photons in superconducting qubits, Science \textbf{358}, 1175 (2017).

%\pacs{03.65.Yz, 42.50.Dv}
\end{thebibliography}
\end{document}